\documentclass[preprint,showkeys,showpacs]{revtex4}
\usepackage{amssymb}
\usepackage{amsmath}
\usepackage{graphicx}
\usepackage{xcolor}
\renewcommand{\figurename}{Fig.}

\begin{document} 
\title{Influence of surface effects on neutron skin in atomic nuclei}
\author{S.V. Lukyanov and A.I. Sanzhur}
\affiliation{Institute for Nuclear Research, 03680 Kyiv, Ukraine}
\keywords{neutron skin, Gibbs-Tolman approach, direct variational method, Skyrme forces}
\pacs{21.10.Gv, 13.75.Cs, 21.60.Ev}

\begin{abstract}
The influence of the diffuse surface layer of a finite nucleus on the mean square
radii and their isotopic shift is investigated. We present the calculations
within the Gibbs-Tolman approach using the experimental values of the nucleon
separation energies. Results are compared with that obtained by means of direct
variational method based on Fermi-like trial functions.
\end{abstract}

\maketitle

\section{Introduction} 

Finite nucleus  possesses the surface diffuse layer due to the quantum penetration of particles 
into the classically forbidden region. As a result, there is ambiguity in determination of the 
nuclear size \cite{my.npa.73}. Information on the size of atomic nuclei and average characteristics 
of radial nucleon distributions can be obtained from the values of mean square radii of nuclei
\cite{ba.b.89}. In analysis of experimental data the two-parameter Fermi function is often used 
for the spatial distribution of nucleons, 
\begin{equation}
F_q(r)=\left[1+\exp \left( \frac{r-R_{q}}{a_{q}}\right) \right]^{-1},
\label{ferm}
\end{equation}
where $R_q$ is the half-density radius and $a_q$ is the diffuseness parameter
of the distribution.
Here, $q=n$ is for neutron and $q=p$ for proton distribution. 
For two-component system two different patterns may arise
\cite{waviroce.prc.10} depending on the parameter values $R_q$ and $a_q$. 
For the same values of the diffuseness, $a_n=a_p$, but for different values of radii,
$R_n>R_p$, one considers neutron skin.
In the opposite case of the same values of the radii, $R_n=R_p$, and for
different values of the  diffuseness parameter, $a_n>a_p$, there is a neutron halo.
  Studies show the mixture of two mentioned patterns with approximately equal
contributions \cite{lusa.npae.16}.

  In this paper the effect of the diffuse surface layer of a nucleus on the mean
square radii is considered making the comparison between results of two models.
  First, we adopt the spatial distribution of nucleons having sharp boundary at
the equimolar radius.
  Within the Gibbs-Tolman (GT) approach \cite{gibb28,tolm49,rowi82,%
kolusa.prc86.12} the values of the equimolar radius and the bulk nucleon density
are obtained using the experimental data on nucleon separation energies. 
  Second, we consider the diffuse spatial distribution of nucleons in a nucleus.
  We apply the the direct variational method based on the specific Fermi-like
trial functions \cite{lusa.npae.16,kolusa.prc.12,kolusash.prc.17} and
the bulk nucleon density is normalized to that obtained using the GT approach.
  The comparison of the two above mentioned considerations allow us to allocate
the effects of the surface layer on rms radii and their isotopic shift.
  Sec.~\ref{sec:model} gives the basics of the Gibbs-Tolman approach and direct
variational method.
  Results and discussion are presented in Sec.~\ref{sec:calc}, conclusions are
summarized in Sec.~\ref{sec:concl}.

\section{The model}\label{sec:model}

\subsection{The Gibbs-Tolman approach}

  Following the Gibbs-Tolman approach, we consider the spatial distribution of
nucleons in spherical nucleus having the sharp boundary located within the
surface region.
  The dividing spherical surface of radius $R$ separates the nucleus into bulk
and surface parts with the corresponding volume $V=4\pi R^3/3$ and surface area
$S=4\pi R^2$.
  The total energy $E$ of the nucleus is also divided into the volume, $E_V$,
and, the surface, $E_S$, parts, respectively. Namely,
\begin{equation}
E=E_V + E_S +E_C.
\label{Ne1}
\end{equation}
  Here the Coulomb energy $E_{C}$ is fixed and does not depend on the dividing
radius $R$.
  The volume energy $E_V$ is considered as the energy of homogeneous nuclear 
matter $E_V=E_{\infty}$ contained in the volume $V$.

  We consider the two-component nuclear matter with the neutron-proton asymmetry
parameter $X=(N-Z)/A$, where $N$ and $Z$ are, respectively, the neutron and
proton numbers, $A=N+Z$ is the mass number.
  The neutron, $\mu _{n}$, and proton, $\mu _{p}$, chemical potentials are
defined as
\begin{equation}
\mu _{n}=\left. \frac{\partial E_V}{\partial N}\right\vert_Z\ , 
\quad 
\mu _{p}=\left. \frac{\partial E_V}{\partial Z}\right\vert_N\ .  
\label{demuq}
\end{equation}
  By the assumption of the GT approach, the nuclear matter inside certain volume
is taken to be in a state having the same values of chemical potentials as those
of real nucleus (see~\cite{kolusash.prc.17})
\begin{equation}
\mu_q(\{\rho_{q,V}\})=-s_q -\lambda_{q,C}, 
\label{muq}
\end{equation}
where $s_q$ is the single-nucleon separation energy, $\rho_{q,V}$ is the bulk
nuclear matter density of the step $r$-distribution
\begin{equation}
\rho_q(r)=\rho_{q,V}\ \Theta(R_{s,q}-r),
\label{stepf}
\end{equation}
where $R_{s,q}$ are the partial (neutron and/or proton) radii.
  The Coulomb contribution $\lambda_{q,C}$ to the chemical potential
$\lambda_{q}=-s_q$ of the nucleus is subtracted in (\ref{muq}) since the
resulting value $\mu_ {q}$ of Eq.~(\ref{demuq}) corresponds to uncharged
nuclear matter.
  The value of Coulomb contribution in (\ref{muq}) is determined by
\begin{equation}
\lambda_{n,C}=\left. \frac{\partial E_{C}}{\partial N}\right\vert_{Z},
\quad 
\lambda_{p,C}=\left. \frac{\partial E_{C}}{\partial Z}\right\vert _{N}.  
\label{lambdaC}
\end{equation}
   Below we will approximate the Coulomb energy $E_C(X)$ by the smooth function
\begin{equation}
E_C(X) = e_C(A) (1-X)^2 A,
\end{equation}
where
\[
e_C(A)= 0.207 A^{2/3} - 0.174 A^{1/3}
\]
is the Coulomb energy parameter estimated from the fit to the experimental data,
see~\cite{kosa.ijmpe.13}.

Considering the asymmetric nuclear matter with the asymmetry parameter $X\ll 1$, 
the bulk energy per particle can be used as \cite{kolusash.prc.17} 
\begin{equation}
E_{V}/A\equiv e_{0}(\rho_{V})
+e_{2}(\rho_V)\left( \frac{\rho_{-,V}}{\rho_V}\right)^{2}, 
\label{Ebulk}
\end{equation}
where
\begin{equation}
e_{0}(\rho_{V})=\frac{\hbar^{2}}{2m}\alpha \rho_{V}^{2/3}
+\frac{3t_{0}}{8}\rho_{V} 
+\frac{t_{3}}{16}\rho_{V}^{\nu +1}
+\frac{\alpha }{16}\left[ 3t_{1}+t_{2}(5+4x_{2})\right] \rho_{V}^{5/3} 
\label{ee0}
\end{equation}
and
\begin{equation}
e_{2}(\rho_V)=\frac{5}{9}\frac{\hbar ^{2}}{2m}\alpha \rho_{V}^{2/3}
-\frac{t_{0}}{8}(1+2x_{0})\rho_{V} 
-\frac{t_{3}}{48}(1+2x_{3})\rho_{V}^{\nu +1}
+\frac{5\alpha }{72}\left( t_{2}(4+5x_{2})-3t_{1}x_{1}\right) \rho_{V}^{5/3}.   
\label{ee2}
\end{equation}
  Here $\alpha={(3/5)}\,(3\,\pi^2/2)^{2/3}$, $\rho_V=\rho_{n,V}+\rho_{p,V}$ and 
$\rho_{-,V}=\rho_{n,V}-\rho_{p,V}$ are the total nucleon and the neutron excess
bulk densities, respectively, $t_{i}$, $x_{i}$ and $\nu$ are the parameters of 
Skyrme force.

The surface energy is given by \citep{kolusash.prc.17}
\begin{equation}
E_S=(\sigma+\mu\rho_S+\mu_{-}\rho_{-,S})S,
\label{ES}
\end{equation}
where $\sigma$ is the surface tension coefficient. Here $\mu=(\mu_n+\mu_p)/2$ 
and $\mu_{-}=(\mu_n-\mu_p)/2$ are, respectively, the isoscalar and isovector chemical potentials, 
$\rho_S=\rho_{n,S}+\rho_{p,S}$ is the surface density and $\rho_{-,S}=\rho_{n,S}-\rho_{p,S}$ 
is the isovector surface density (see details in Ref. \citep{kolusash.prc.17}).

In accordance with the GT concept, the actual equimolar radius $R_e$ of the droplet is determined by the
requirement that the contribution to $E_S$ from the bulk term
in Eq. (\ref{ES}) should be excluded. This requirement can be satisfied if the following condition
is fulfilled:
\begin{equation}
\left(\mu\rho_S+\mu_{-}\rho_{-,S}\right)_{R=R_e}=0.
\label{cond_Re}
\end{equation}
Eq. (\ref{cond_Re}) determines the equimolar radius $R_e$.

  As soon as the chemical potentials of a nucleus is known, one obtains the
partial volume densities $\rho_{q,V}$ using Eqs.~(\ref{demuq}) -- (\ref{ee2}).
  Then, calculating the partial surface densities
\begin{equation}
\rho_{n,S}[R]=\frac{N}{4\pi R^{2}}-\frac{1}{3}\rho_{n,V}R\
,\quad 
\rho _{p,S}[R]=\frac{Z}{4\pi R^{2}}-\frac{1}{3}\rho_{p,V}R
\label{rhoPART}
\end{equation}
and applying the condition (see also Eq. (\ref{cond_Re}))
\begin{equation}
\rho _{q,S}[R_{q,e}]=0,
\end{equation} 
one finds the partial equimolar radii $R_{q,e}$ \cite{kolusash.prc.17}.
The root mean square (rms) radius for the nucleon density distribution
$\rho_{q}{(\mathbf{r})}$ is defined as
\begin{equation}
\sqrt{\left\langle r_{q}^{2}\right\rangle}
=\sqrt{\left. \int {d\mathbf{r}\,r^{2}\,\rho}_{q}{(\mathbf{r})}\right/ 
\int {d\mathbf{r}\,\,\rho}_{q}{(\mathbf{r})}}\ .
\label{rmsrq}
\end{equation}
In particular, for the step distribution function (\ref{stepf}), the
corresponding rms radii are given by
\begin{equation}
\sqrt{\left\langle r_{q}^{2}\right\rangle}=\sqrt{\frac{3}{5}}R_{q,e}.
\label{rmsreq}
\end{equation}

\subsection{The direct variational method}

In order to consider the asymmetry of the diffuse surface of the spatial distribution of nucleons, 
according to the direct variational method (see, for example, \cite{kosa.epj.08, kolusash.prc.17}), 
we adopt the trial function for $\rho_{q}(\mathbf{r})$ as a power of the Fermi function, namely
\begin{equation}
\rho_{q}(\mathbf{r})=\rho_{0,q} F_q(r)^{\xi_q},
\label{fermxi}
\end{equation}
where $\rho_{0,q}$, $R_{q}$, $a_{q}$ and $\xi_q$ are the variational parameters.
The profile function $\rho_{q}(r)$ should satisfy the conservation conditions
for numbers of neutrons and protons
\begin{equation}
\int d\mathbf{r} \rho_n(\mathbf{r}) = N, 
\quad
\int d\mathbf{r} \rho_p(\mathbf{r}) = Z.
\label{nzcons}
\end{equation}

The total energy of a nucleus is given by
\begin{equation}
E_{\mathrm{tot}}=E_{\mathrm{kin}}+E_{\mathrm{Sk}}+E_C,
\end{equation}
where $E_\mathrm{kin}$ is the kinetic energy, $E_\mathrm{Sk}$ is the potential energy of the Skyrme interaction, 
and $E_C$ is the Coulomb energy. In the case of finite nuclei the kinetic energy is
\begin{equation}
E_\mathrm{kin}=\int d\mathbf{r}\ \epsilon_\mathrm{kin}(\mathbf{r}),
\end{equation}
where the kinetic energy density $\epsilon_\mathrm{kin}(\mathbf{r})$ is given by the sum of the neutron 
and proton contributions
\begin{equation}
\epsilon_\mathrm{kin}(\mathbf{r})=\epsilon_\mathrm{kin,n}(\mathbf{r})+\epsilon_\mathrm{kin,p}(\mathbf{r}).
\label{ekinsum}
\end{equation}
We adopt the extended Thomas-Fermi approximation for the kinetic-energy density \cite{brguha.pr.85}
\begin{equation}
\epsilon_\mathrm{kin,q}(\mathbf{r})
={\frac{\hbar^{2}}{2m}}\left[ {\frac{3}{5}}\,(3\,\pi ^{2})^{2/3}\,\rho_{q}^{5/3}
+ \frac{1}{36}{\frac{(\mathbf{\nabla }\rho_{q})^{2}}{\rho_{q}}}+{\frac{1}{3}}\,\nabla^{2}\rho_{q}\right].
\label{eqkin}
\end{equation}
In our consideration, the potential energy $E_\mathrm{Sk}$ also includes gradient terms due to the spin-orbit 
interaction. We note that pair interactions are not considered here.

For the ground state of the nucleus, the values of the variational parameters
can be found by minimizing the total energy of the nucleus with respect to all
possible small changes of the variational parameters, provided the conditions
(\ref{nzcons}) are fulfilled. Below, in the subsequent calculations, the values 
of the nucleon density parameters $\rho_{0,q}$ will be normalized to the values
obtained within the GT approach using the experimental data on the chemical potentials, 
see also Eq. (\ref{muq}),
\begin{equation}
\rho_{0,q}=\rho_{q,V}.
\label{rhonorm}
\end{equation}
  In view of Eqs.~(\ref{fermxi}) and  (\ref{nzcons}) the conditions for the particle
number conservation are written by
\begin{equation}
\int d\mathbf{r} F_n(r)^{\xi_n} = \frac{N}{\rho_{n,V}}\ , 
\quad
\int d\mathbf{r} F_p(r)^{\xi_p} = \frac{Z}{\rho_{p,V}}\ .
\label{nzconsnew}
\end{equation}
  Thus, fixing the values of $\rho_{0,q}$ and $R_{q}$ using the relations
(\ref{rhonorm}) and (\ref{nzconsnew}) the number of free variational parameters is
reduced to four, that are $a_{q}$ and $\xi_q$.
  For the trial functions are given by (\ref{fermxi}) one can obtain the
leptodermous expansion ($a_q/R_q\ll 1$) of the rms radius \cite{lusa.npae.16}:
\begin{eqnarray}
\sqrt{\left\langle r^2 \right\rangle_q} \simeq
\sqrt{\frac{3}{5}} R_q \left\{1+\kappa_0(\xi_q)\frac{a_q}{R_q}
-\frac{7}{2}\left(\kappa_0^2(\xi_q)
-2\kappa_1(\xi_q)\right)\left(\frac{a_q}{R_q}\right)^2 \right. \nonumber \\
+\left. \frac{1}{6}\left(75\kappa_0^3(\xi_q)-204\kappa_0(\xi_q)
\kappa_1(\xi_q)+81\kappa_2(\xi_q)\right)
\left(\frac{a_q}{R_q}\right)^3 \right\}\ , 
\label{rmsrq_3}
\end{eqnarray}
where the coefficients $\kappa_j(\xi)$ are the generalized Fermi integrals,
\begin{equation}
\kappa_j(\xi )=\int_0^\infty {dx\,x^j\left[ (1+e^x)^{-\xi}-
(-1)^j\left(1-(1+e^{-x})^{-\xi}\right)\right]}\ .
\label{ki}
\end{equation}

\section{Numerical calculations}\label{sec:calc}

\begin{figure}
\begin{center}
\includegraphics*[scale=0.6,clip]{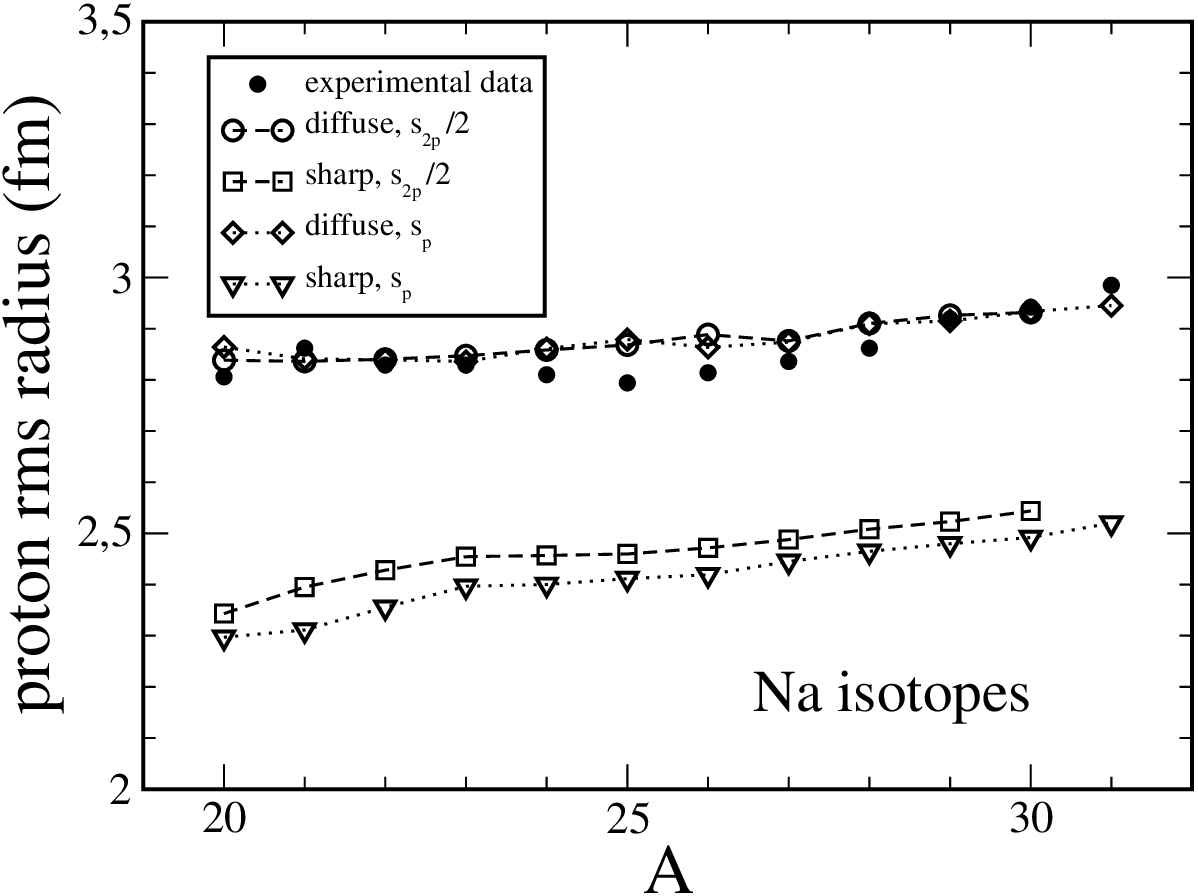}
\end{center}
\caption{Dependence of the effective rms radii for proton spatial distributions in 
Na isotopes on mass number $A$. The black circles are the experimental data \cite{sugebo.prl.95},
the triangles are the calculations within the GT approach using one-proton separation energy;
the rhombuses are the calculations for diffuse distribution with one-proton separation energy;
the squares are the calculations in the framework of the GT approach, half of the two-proton 
separation energy value was used; the circles are the calculations for diffuse distribution 
with half of the two-proton separation energy value.}
\label{fig1}
\end{figure}

Here we present the results of numerical calculations for the neutron and proton rms radii for isotopes of sodium, 
tin, and lead. The SkM$^\ast$ parameterization \cite{brguha.pr.85} for the Skyrme nucleon-nucleon interaction was 
used in the calculations. Since sodium isotopes $^{21-24,28-31}$Na have the observed prolate deformation 
\cite{sugebo.prl.95}, then we will consider the effective rms radii. \figurename~\ref{fig1} shows the calculation 
results for the effective rms radii of the proton spatial distributions in sodium isotopes versus the mass
number $A$. 
  As the charge number is fixed, the figure actually depicts the dependence
on the number of neutrons $N=A-11$.
  The triangles indicate the calculation in the framework of the GT approach for
the sharp distribution (\ref{stepf}) according to the formula (\ref{rmsreq}).
The rhombuses indicate the calculation using the direct variational method for
the diffuse distribution (\ref{fermxi}) in accordance with the expression (\ref{rmsrq_3}).
In both cases, calculations were done using the experimental values of the
one-proton separation energy $s_p$ \cite{AuWaWaKoMcPf.cpc.12} for the proton
chemical potential $\mu_p$ in accordance with (\ref{muq}). 
  For clarity, the points are connected by dotted lines.
  As can be seen from the figure, the triangles are located much lower than
the experimental data, while the rhombuses are almost identical with them.
  The difference between the upper and lower graphs is about of $0.5$ fm.
  So, the account of the diffuse edge in spatial distribution of protons
increases the proton rms radii by about of $20\%$. 
The results of the calculation with half values of the two-particle nucleus
separation energy $s_{2n}/2$ and $s_{2p}/2$ almost coincide with the calculations 
for single-nucleon separation energies. Here and below, we did not perform calculations
for isotopes with no experimental data are available.

\begin{figure}
\begin{center}
\includegraphics*[scale=0.6,clip]{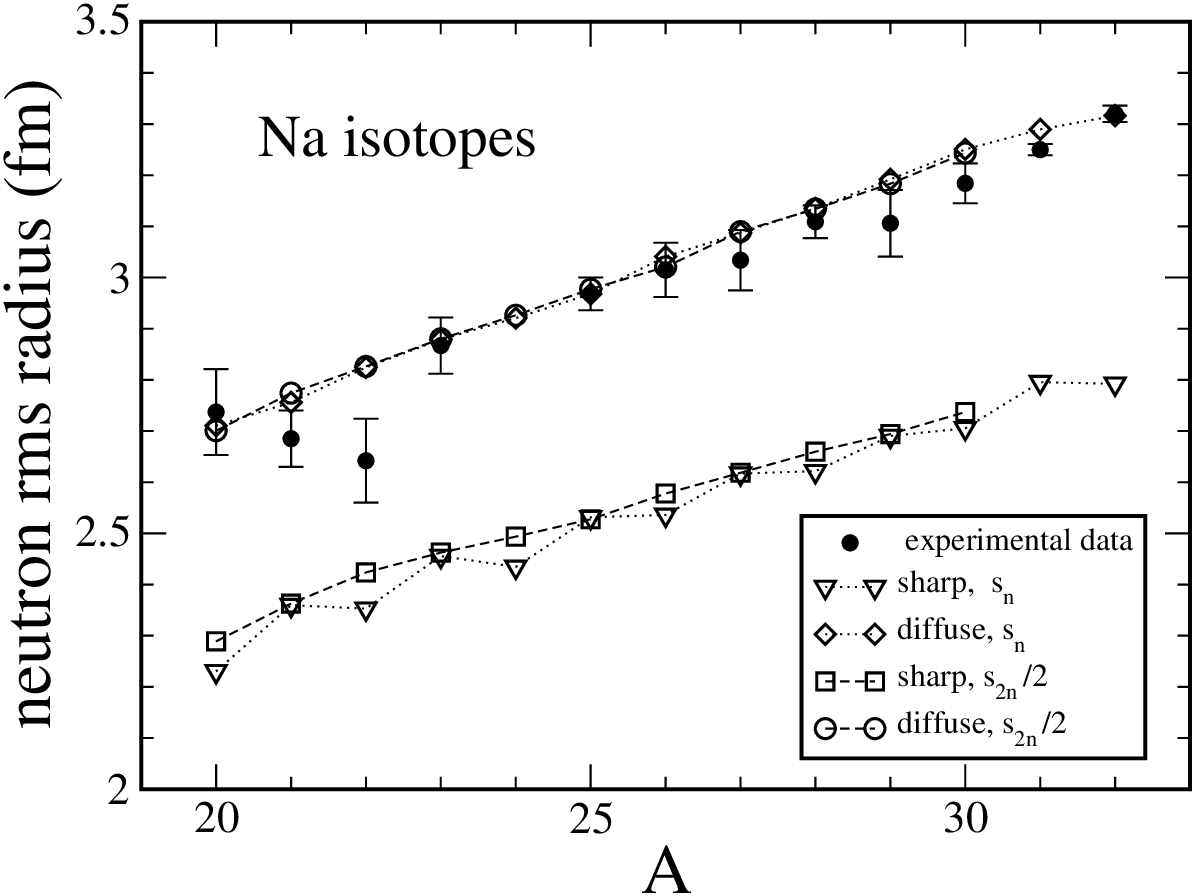}
\end{center}
\caption{Dependence of the effective rms radii for neutron spatial distributions in sodium 
isotopes on mass number $A$. The black circles are the experimental data \cite{sugebo.prl.95}, 
the triangles are the calculations in the framework of the GT approach using one-neutron 
separation energy for neutron chemical potential; the squares are the calculations in the 
framework of the GT approach with half the two-neutron separation energy value;
the rhombuses are the calculations for diffuse distribution with one-neutron separation energy; 
the circles are the calculations for diffuse distribution with half the two-neutron separation 
energy value.}
\label{fig2}
\end{figure}

\figurename~\ref{fig2} shows the results of calculations of the neutron effective
rms radii in sodium isotopes as a function of mass number $A$ together with the
experimental data.
  For notations similar to those of \figurename~\ref{fig1} calculations were
done using the one-neutron separation energy $s_n$ \cite{AuWaWaKoMcPf.cpc.12}
for the neutron chemical potential $\mu_n$ in accordance with (\ref{muq}).
  It is seen from \figurename~\ref{fig2} that the triangles are located below
the experimental data by about of $0.5$ fm.
  The rhombuses reproduce the experimental data fairly well and show the
monotonous growth as mass the number increases.
  One should notice the sawtooth behavior for the calculation marked by
triangles.
  This calculation corresponds to the one-neutron separation energy $s_n$ taken
for the neutron chemical potential.
  The sawtooth behavior disappears and the $A$-dependence of neutron rms radius
becomes monotonous if we use the half-value of the two-neutron separation
energy, $s_{2n}/2$, for the neutron chemical potential, see squares connected by
the dashed line in \figurename~\ref{fig2}.
  Such sawtooth dependence is a manifestation of the pairing effect which
contributes to the single neutron separation energy $s_n$ and cancels out in
$s_{2n}/2$.
  The pairing effect is not that pronounced if the diffuse neutron distribution is
used, the calculations using $s_n$ (circles) and $s_{2n}/2$ (rhombuses) for
the neutron chemical potential are practically coincide,
see \figurename~\ref{fig2}.
It should be noted that in our model the pairing effects are manifested 
exclusively throuhg the experimental values of one-particle neutron and proton chemical potentials.
  We note, that the use of experimental values of $s_q$ for chemical potentials
still does not allow to reproduce well the fine structure of the mass number
dependence of measured rms radii.

\begin{figure}
\begin{center}
\includegraphics*[scale=0.6,clip]{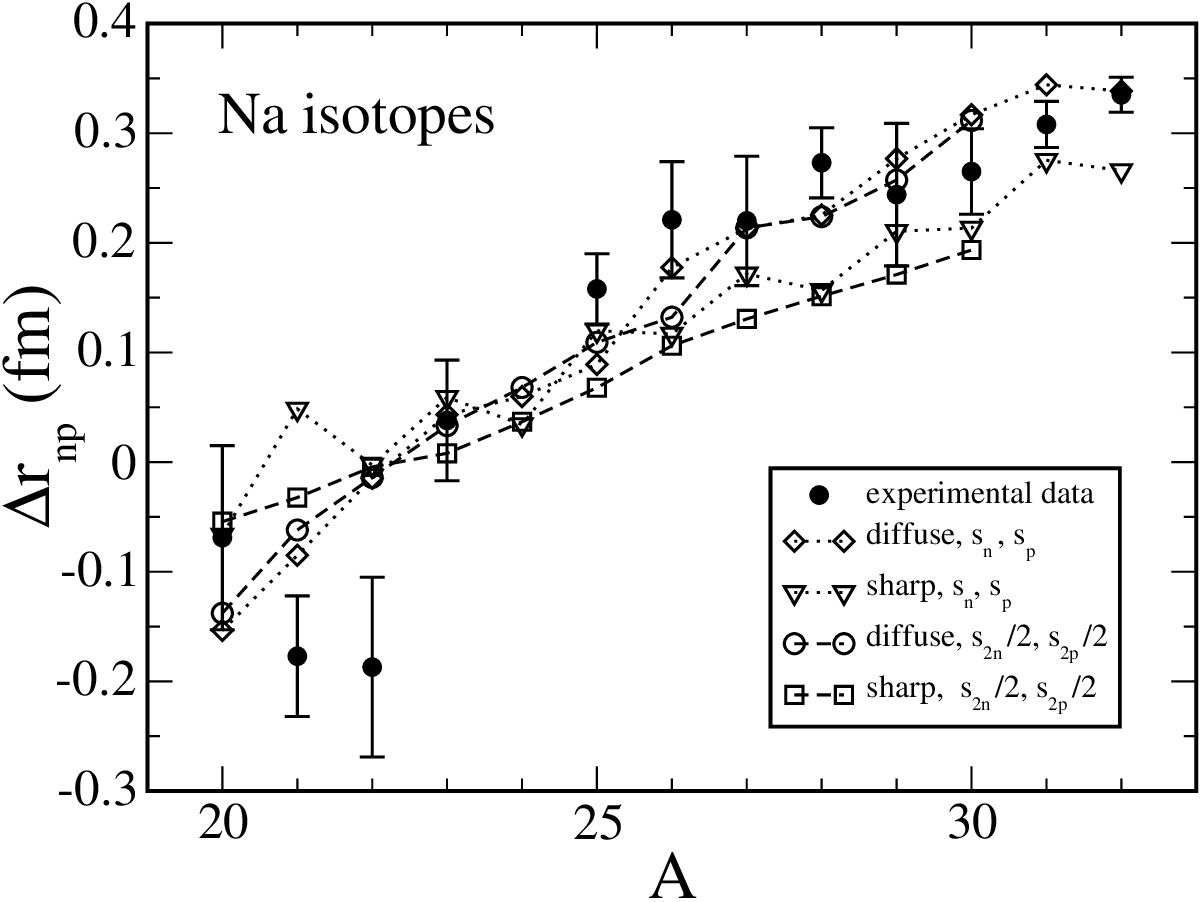}
\end{center}
\caption{Difference $\Delta r_{np}$ between the effective rms radii of the neutron and proton 
spatial distributions as a function of $A$ for Na isotopes. The notations are the same as 
in Figs. \ref{fig1} and \ref{fig2}.}
\label{fig3}
\end{figure}

\begin{figure}
\begin{center}
\includegraphics*[scale=0.6,clip]{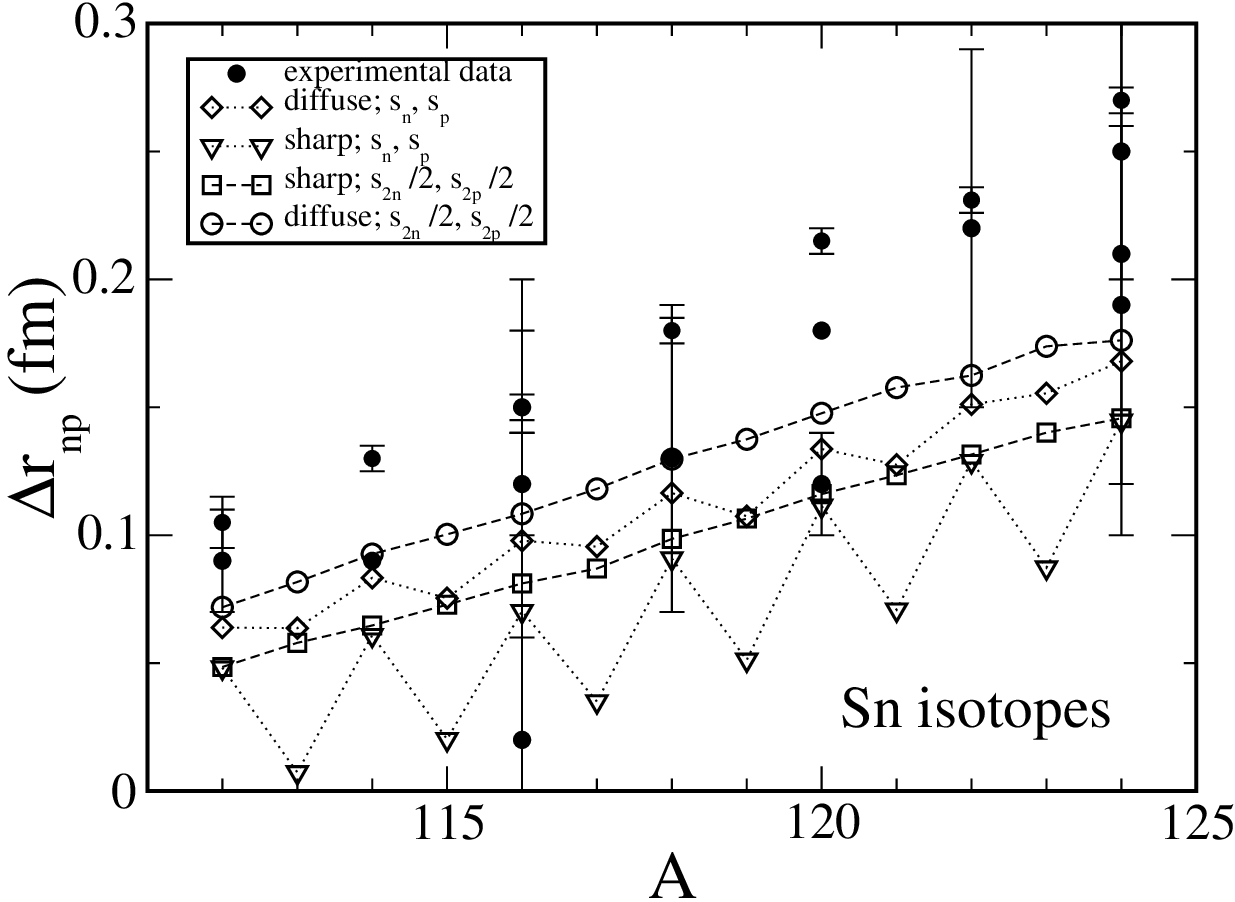}
\end{center}
\caption{Difference $\Delta r_{np}$ between the rms radii of
the neutron and proton spatial distributions as a function of $A$ for Sn
isotopes. The notations are the same as in \figurename~\ref{fig3}.
The experimental data were taken from Refs. \cite{ray.prc.79,trjalu.prl.01,krak.np.04}.}
\label{fig4}
\end{figure}
\begin{figure}
\begin{center}
\includegraphics*[scale=0.6,clip]{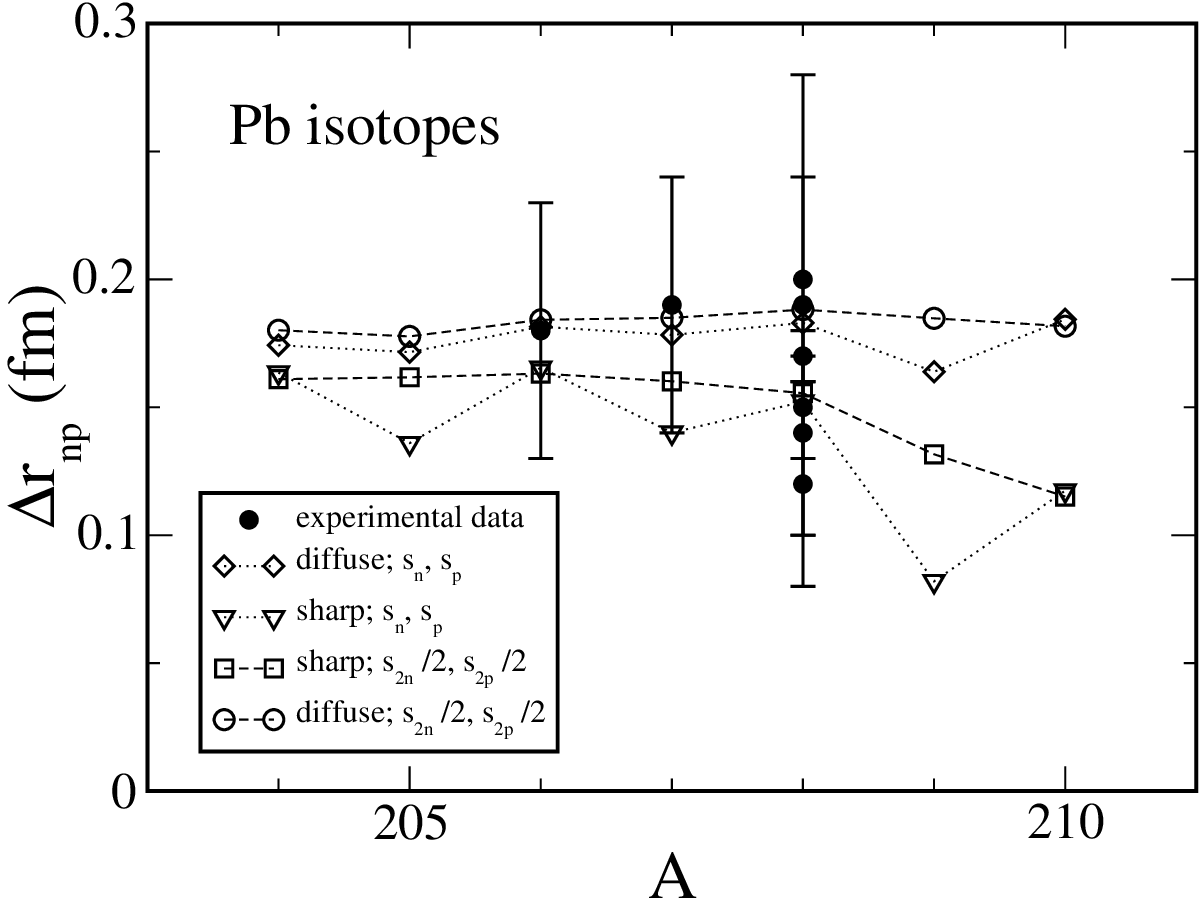}
\end{center}
\caption{Difference $\Delta r_{np}$ between the rms radii of the
neutron and proton spatial distributions as a function of $A$ for Pb isotopes.
The notations are the same as in \figurename~\ref{fig3}.
The experimental data were taken from Refs. \cite{sth.prc.94,kaambrde.prc.02,clkeha.prc.03}.}
\label{fig5}
\end{figure}

Figs.~\ref{fig3}, \figurename~\ref{fig4}, and \figurename~\ref{fig5} 
depict the calculation results for difference between the neutron and proton rms radii
$$
\Delta r_{np}=\sqrt{\left\langle r^2\right\rangle_n}-
\sqrt{\left\langle r^2\right\rangle_p}
$$
for Na, Sn, and Pb isotopes in comparison with the experimental data \cite{trjalu.prl.01}. 
\figurename~\ref{fig3} shows the difference between \figurename~\ref{fig2} 
and \ref{fig1}. 
As seen from \figurename~\ref{fig3} the experimental data are described quite well with all four
calculations presented. Although Figs. 1 and 2 show that step-like distributions underestimate 
the rms radii by an average of about 20\%, nevertheless, when calculating the difference, 
such shifts are mutually compensated. The diffuse distribution calculations demostrate 
slightly steeper slopes than the stepped distribution calculations. 
This can be explained by the fact that for the diffuse distribution, the rms neutron radii 
increase more rapidly with the increase of the number of neutrons $N$ than for the step-like distribution. 
The fine structure, however, is not reproduced, especially within the region of neutron-deficient isotopes.
In general, the calculations with diffuse distribution are better described the experimental data.

In Figs. \ref{fig4} and \ref{fig5} there is noticeable difference (of about 25 -- 30\%) between
two types of calculations which correspond to the diffuse and stepped nucleon distributions.
  The calculations for the diffuse nucleon distribution (marked as rhombuses)
give a better description of the experimental data and are located higher
than calculations for the stepped nucleon distribution (marked as triangles).
  In both Figs. \ref{fig4} and \ref{fig5} the sawtooth
$A$-dependence is clearly seen for $\Delta r_{np}$ obtained using single
nucleon separation energies $s_q$.
  This indicates the pairing effect contribution to the isotopic difference in
the root mean square radii. 
  The sawtooth dependence is eliminated by the use of the half-value of
the experimental two-nucleon separation energies $s_{2n}/2$ and $s_{2p}/2$ for
the corresponding chemical potentials, see squares and circles in
Figs.~\ref{fig4} and \ref{fig5}.

\begin{figure}
\begin{center}
\includegraphics*[scale=0.6,clip]{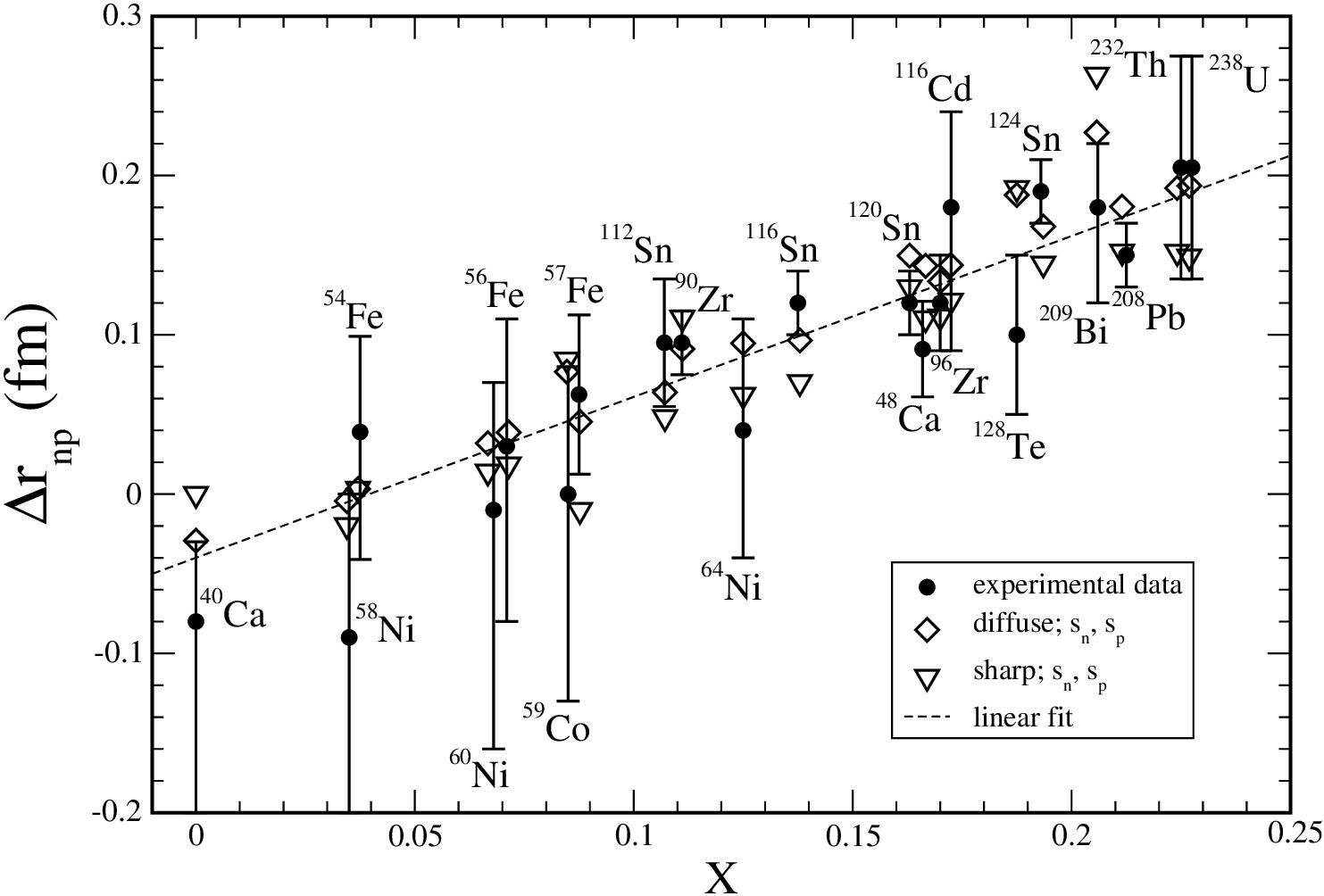}
\end{center}
\caption{Difference $\Delta r_{np}$ between the rms radii of the
neutron and proton spatial distributions as a function of the asymmetry
parameter $X$ for a set of nuclei. The experimental data are taken from
\cite{trjalu.prl.01}, the dashed line is the linear approximation taken from
\cite{trjalu.prl.01}, the triangles are the calculation in the framework of the
GT approach, the rhombuses are the calculation for the diffuse distribution.}
\label{fig6}
\end{figure}

\begin{figure}
\begin{center}
\includegraphics*[scale=0.6,clip]{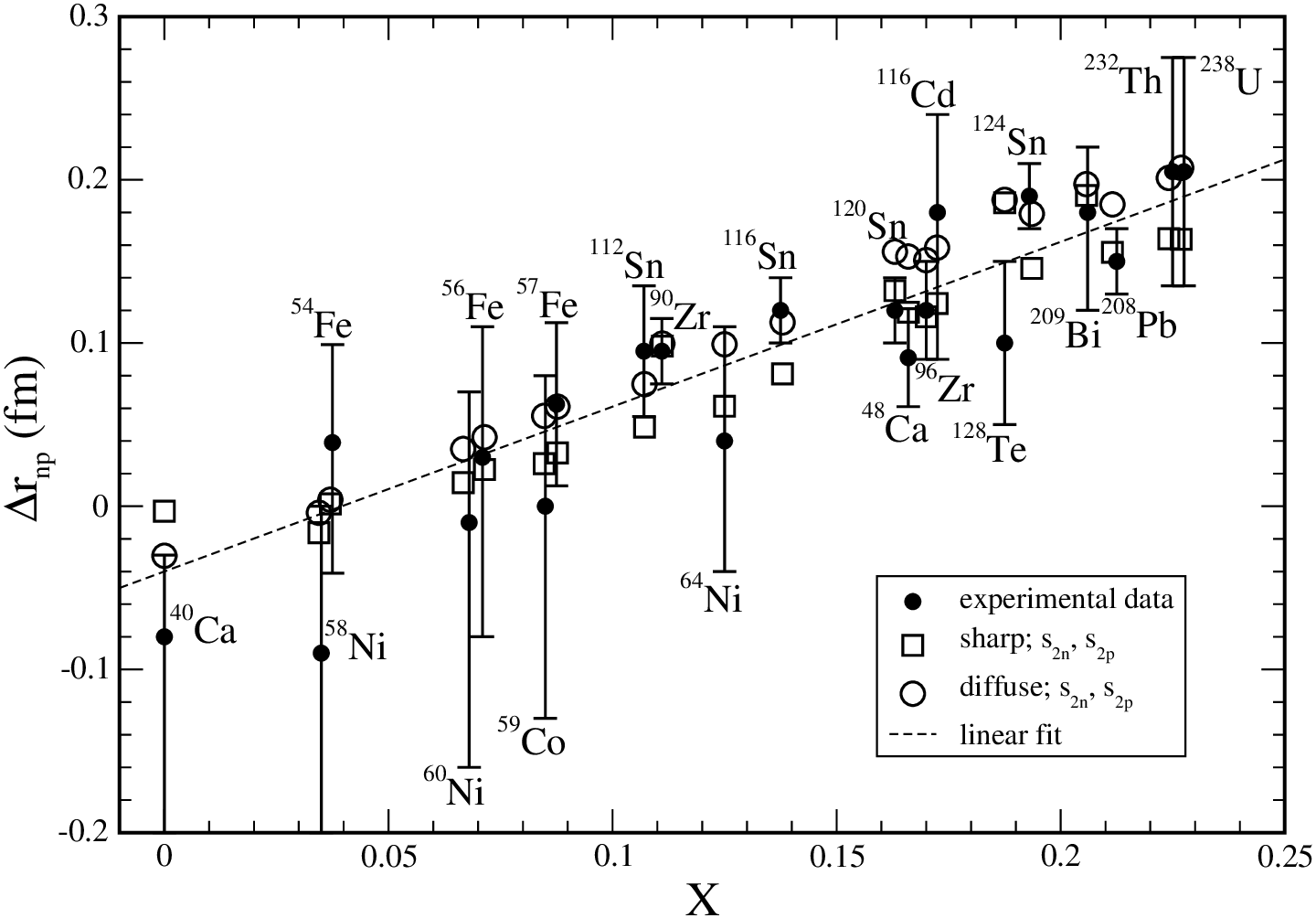}
\end{center}
\caption{Difference $\Delta r_{np}$ between the rms radii of the
neutron and proton spatial distributions as a function of the asymmetry
parameter $X$ for a set of nuclei. The experimental data are taken from
\cite{trjalu.prl.01}, the dashed line is the linear approximation taken
from \cite{trjalu.prl.01}, the squares are the calculation within the GT
approach, the circles are the calculation for the diffuse distribution.}
\label{fig7}
\end{figure}

  The isotopic shift $\Delta r_{np}$ between the neutron and proton rms radii
(neutron skin) is presented in \figurename~\ref{fig6} as a function of the
asymmetry parameter $X$ for different nuclei.
  The experimental data (symbols with error bars) are taken from
\cite{trjalu.prl.01} where the isotopic difference between the rms radii
was estimated as $\Delta r_{np}=(-0.04\pm 0.03)+(1.01\pm 0.15)X$.
  The result of this linear fit is presented by the dashed line in
\figurename~\ref{fig6}.
  Calculations shown in \figurename~\ref{fig6} were performed using one-particle
separation energies for the sharp (triangles) and the diffuse (rhombuses)
nucleon distributions.
  As seen from the figure, both types of calculations are mostly located within
the limits of experimental errors. 
  \figurename~\ref{fig7} shows similar calculations as in 
\figurename~\ref{fig6} except half the values of the two-particle
separation energy are taken for the chemical potentials instead of the
one-particle one to exclude the pairing effect.

  In contrast to the significant shift of about 0.5 fm in proton and neutron rms
radii due to the presence of diffuse layer in spatial nucleon distribution
(compare triangles and rhombuses, respectively, in Figs.~\ref{fig1} and
\ref{fig2}), the contribution from the diffuse layer has an only slight effect on
the isotopic shift $\Delta r_{np}$ as can be concluded from Figs.~\ref{fig6} and
\ref{fig7} paying attention to the location difference between triangles and
rhombuses in \figurename~\ref{fig6} and also between squares and circles
in \figurename~\ref{fig7}.
  The reason for the weak sensitivity of $\Delta r_{np}$ on the presence of the
diffuse layer is that the contributions to rms radii gained from the diffuse
surface are partially canceled in the resulting isotopic difference.
  This jusifies the application of simple nucleon distribution (\ref{stepf}) in
describing the properties of the neutron skin $\Delta r_{np}$.

\section{Conclusions}\label{sec:concl}

  In this paper, we have studied the influence of the diffuse surface of a
nucleus on its rms radii and their difference by comparing the results of
calculations for two cases.
  In the first case, in the framework of the Gibbs-Tolman approach, we
considered the stepped spatial distribution of nucleons having formal equimolar
radius located within the surface region of a nucleus.
The bulk density was determined by adjusting the values of the chemical
potentials to their experimental values using the nuclear matter equation of
state.
  In the second case, the direct variational method was used applying
Fermi-like distribution function for the spatial distribution of nucleons.
The neutron and proton densities in the center of a nucleus were normalized
to the values obtained within the Gibbs-Tolman approach.

  It is found that the use the diffuse nucleon distribution gives a better
description of the experimental rms radii as demonstrated in Figs.~\ref{fig1}
and \ref{fig2} for sodium isotopes.
  The contribution from the diffuse surface layer increases the neutron and
proton rms radii by about 20\% as compared to the stepped nucleon distribution.
For sodium isotopes, the neutron rms radius exhibits a monotonic increase with
increasing mass number as seen from \figurename~\ref{fig2}.

  The isovector shift $\Delta r_{np}$ between the neutron and proton rms radii was 
calculated for sodium, tin, and lead isotopes using both the diffuse and stepped nucleon 
distributions. For sodium, tin, and led isotopes the use of diffuse Fermi-like distribution
allows  better reproduction of the experimental values $\Delta r_{np}$.
The influence of the pairing effect on the isovector shift $\Delta r_{np}$
is demonstrated in Figs~\ref{fig3}, \ref{fig4}, and \ref{fig5} for Na, Sn, and Pb isotopes.
  The sawtooth behavior of $\Delta r_{np}(A)$ reflects the odd-even effect in
the one-nucleon separation energies $s_q$ used for corresponding chemical
potentials.
  After replacing the one-particle separation by the half-value of the
two-particle separation energy the mentioned behavior disappears and
practically monotonic dependence on the mass number is obtained for
$\Delta r_{np}$.
  The calculations of the neutron skin for certain set of nuclei, from light to
heavy masses, depending on the asymmetry parameter, show that both models
describe the experimental data within the experimental errors.

\section{Acknowledgments}

The work was supported in part by the Fundamental Research program "Fundamental
research in high energy physics and nuclear physics (international collaboration)"
at the Department of Nuclear Physics and Energy of the National Academy of Sciences 
of Ukraine. S.V.L. and A.I.S. thank the support in part by the budget program 
“Support for the development of priority areas of scientific researches”, 
the project of the Academy of Sciences of Ukraine, Code 6541230.


\begin{thebibliography}{99}

\bibitem{my.npa.73}
W.D. Myers. 
Geometric properties of leptodermous distributions with applications to nuclei. 
Nucl. Phys. A 204 (1973) 465.

\bibitem{ba.b.89}
C.J. Batty et al. 
Experimental Methods for Studying Nuclear Density Distributions. 
In: Advances in Nuclear Physics. Ed. by J.W. Negele and E. Vogt
(New York: Plenum Press, 1989) Vol. 19, p. 1.

\bibitem{waviroce.prc.10}
M. Warda et al. 
Analysis of bulk and surface contributions in the neutron skin of nuclei. 
Phys. Rev. C 81 (2010) 054309.

\bibitem{lusa.npae.16}
S.V. Lukyanov, A.I. Sanzhur. 
Neutron skin and halo in medium and heavy nuclei within the extended Thomas - Fermi theory. 
Nucl. Phys. At. Energy 17(1) (2016) 5.

\bibitem{gibb28} 
J.W. Gibbs. 
In: The Collected Works 
(New York: Longmans, Green and Co., 1928) Vol. I, p. 219.

\bibitem{tolm49} 
R.C. Tolman. 
The effect of droplet size on surface tension. 
J. Chem. Phys. 17(3) (1949) 333.

\bibitem{rowi82} 
J.S. Rowlinson, B. Widom. 
Molecular Theory of Capillarity 
(Oxford: Clarendon Press, 1982).

\bibitem{kolusa.prc86.12}
V.M. Kolomietz, S.V. Lukyanov, A.I. Sanzhur.
Curved and diffuse interface effects on the nuclear surface tension. 
Phys. Rev. C 86 (2012) 024304.

\bibitem{kolusash.prc.17}
V.M. Kolomietz, S.V. Lukyanov, A.I. Sanzhur, and S. Shlomo. 
Equation of state and radii of finite nuclei in the presence of a diffuse surface layer.
Phys. Rev. C 95 (2017) 064305.

\bibitem{kolusa.prc.12}
V.M. Kolomietz, S.V. Lukyanov, A.I. Sanzhur. 
Nucleon distribution in nuclei beyond the $\beta$-stability line. 
Phys. Rev. C 85 (2012) 034309.

\bibitem{kosa.ijmpe.13}
V.M. Kolomietz, A.I. Sanzhur. 
Thin structure of $\beta$-stability line and symmetry energy. 
Int. Jour. Mod. Phys. E 22(1) (2013) 1350003.

\bibitem{kosa.epj.08}
V.M. Kolomietz, A.I. Sanzhur. 
Equation of state and symmetry energy within the stability valley. 
Eur. Phys. J. A 38 (2008) 345.

\bibitem{brguha.pr.85}
M. Brack, C. Guet, H.-B. H\aa kansson. 
Selfconsistent semiclassical description of average nuclear properties 
– a link between microscopic and macroscopic models. 
Phys. Rep. 123 (1985) 275.

\bibitem{sugebo.prl.95}
T. Suzuki et al. 
Neutron skin in Na isotopes studied via their interaction cross sections. 
Phys. Rev. Lett. 75 (1995) 3241.

\bibitem{AuWaWaKoMcPf.cpc.12}
G. Audi et al. 
The Ame2012 atomic mass evaluation. (I). Evaluation of input data, adjustment procedures.
Chin. Phys. C 36(12) (2012) 1287; 
M. Wang et al.
The Ame2012 atomic mass evaluation. (II). Tables, graphs, and references. 
Chin. Phys. C 36(12) (2012) 1603.

\bibitem{trjalu.prl.01}
A. Trzci\'{n}ska et al. 
Neutron density distributions deduced from antiprotonic atoms. 
Phys. Rev. Lett. 87 (2001) 082501.

\bibitem{ray.prc.79}
L. Ray. 
Neutron isotopic density differences deduced from 0.8 GeV polarized proton elastic scattering.
Phys. Rev. C 19 (1979) 1855.

\bibitem{krak.np.04} 
A. Krasznahorskay et al. 
Neutron-skin thickness in neutron-rich isotopes. 
Nucl. Phys. A 731 (2004) 224.

\bibitem{sth.prc.94}
V.E. Starodubsky, N.M. Hintz. 
Extraction of neutron densities from elastic proton scattering by 206,207,208 Pb at 650 MeV. 
Phys. Rev. C 49 (1994) 2118.

\bibitem{kaambrde.prc.02}
S. Karataglidis et al. 
Discerning the neutron density distribution of 208 Pb from nucleon elastic scattering.
Phys. Rev. C 65 (2002) 044306.

\bibitem{clkeha.prc.03}
B.C. Clark, L.J. Kerr, S. Hama. 
Neutron densities from a global analysis of medium-energy proton-nucleus elastic scattering. 
Phys. Rev. C 67 (2003) 054605.
\end{thebibliography}
\end{document}